\documentclass[aps,prl,superscriptaddress,showpacs,amsmath,twocolumn]{revtex4}

\usepackage{amsmath}
\usepackage{amssymb}
\usepackage{graphicx}

\begin{document}


\title{Relation between Zitterbewegung and the charge conductivity, Berry curvature
and the Chern number of multi band systems}

\author{ J{\'o}zsef Cserti
}
\affiliation{Department of Physics of Complex Systems,
E{\"o}tv{\"o}s University,
}

\author{Gyula D\'avid
}
\affiliation{Department of Atomic Physics,
E{\"o}tv{\"o}s University,
H-1117 Budapest, P\'azm\'any P{\'e}ter s{\'e}t\'any 1/A, Hungary}



\begin{abstract}

We show that the charge conductivity for impurity free multi band electronic systems can be expressed in terms of the diagonal and non-diagonal elements of the Zitterbewegung amplitudes while the Berry curvature and the Chern number is related only to the diagonal elements.
Thus, the phenomenon of the Zitterbewegung can no longer be viewed just as an interesting consequence of quantum physics but it has also an experimental relevance.
Moreover, through several examples we demonstrate how efficient our approach is in the analytical calculation of the charge conductivity.

\end{abstract}


\maketitle


\emph{Introduction.}---The Zitterbewegung predicted originally by Schr\"odinger for Dirac electron is a `trembling'
or in other words a rapid oscillatory motion of the center of the free wave packet for relativistic electron~\cite{Schrodinger:cikk}.
Most recently, Schliemann~\emph{et~al.}~\cite{Schliemann_Loss_PhysRevLett.94.206801} predicted
the Zitterbewegung in spintronic systems where the experimental observation of the effect is more realistic
due to the much smaller frequency of the oscillatory motion.
Thus,  the Zitterbewegung can, in principle, be observed not only in the relativistic regime
but for spintronic systems~\cite{Schliemann_Loss_PhysRevLett.94.206801}
and graphene~\cite{Cserti_Zitter-ISI:000242409000014} as well.
This seminal paper has initiated many other works with an aim to demonstrate the appearance of the Zitterbewegung
not only for Dirac electrons but for quasi-particles in condensed matter physics
(see references in our recent work on the general theory of Zitterbewegung~\cite{Cserti_PhysRevB.81.121417},
where alternative expressions for the Zitterbewegung amplitudes and explicit	
forms of the position operator for several systems are presented).

In connection with graphene Katsnelson has already pointed out that the Zitterbewegung resulting in
an oscillating term in the current operator is responsible for the nontrivial behavior of the conductivity
at zero temperature and zero chemical potential~\cite{Katsnelson_Zitter_min-cond:cikk}.
In this paper we show generally that the charge conductivity for a impurity free multi band system is related to
Zitterbewegung.
Such a relation is expected since in the Kubo formula for the charge conductivity one needs to calculate
the velocity operator in Heisenberg picture which includes an oscillatory term due to the Zitterbewegung in the case of multi band systems.
Moreover, we clarify the relationship between Zitterbewegung and the Hall conductivity of insulators with non-interacting Bloch electrons~\cite{Murakami_PhysRevB.69.235206,PhysRevB.78.195424}.

To show the subtle relation between the Zitterbewegung and the charge conductivity we start with a multi band system
described by the most general matrix Hamiltonian in a Bloch wavefunction basis:
$H_{ab}(\textbf{k})$, where $a,b= 1,2, \cdots N$ are the band indices (here $N$ is the number of bands of the system).
Here each matrix element $H_{ab}(\textbf{k})$ is a differentiable function of the wave number $\textbf{k}$ corresponding to the Bloch states.
In Ref.~\cite{Cserti_PhysRevB.81.121417} we calculated the time dependence of the position operator $\textbf{x}(t)= e^{\frac{i}{\hbar}\, H t} \, \textbf{x}(0)\, e^{-\frac{i}{\hbar}\,  H t}$ of the quasi-particle
by decomposing the Hamiltonian into a sum of projection operators: $H  = \sum_a E_a(\textbf{k}) Q_a(\textbf{k}) $, where $E_a(\textbf{k})$ are the distinct eigenvalues of the Hamiltonian at a given wave number $\textbf{k}$,
and $Q_a(\textbf{k})$ are projection operators ($N \times N$ matrix operator) satisfying the usual relations:
$Q_a Q_b = \delta_{a b} \, Q_a $ and $\sum_a Q_a  = I_N$, where $I_N$ is the $N \times N$ unit matrix.
Note that $a=1,2,\cdots, s$, where $s\leq N$ (for degenerate case $s< N$).
We found that the time dependence of the position operator $\textbf{x}(t)$ in Heisenberg picture becomes
\begin{subequations}
\begin{eqnarray}
\label{x_op-Q-main-shift:eq}
 \textbf{x}(t) &=&  \textbf{x}(0) + \textbf{W} t +
 \sum_{a,b}\, \textbf{Z}^{ab}\, \left( e^{i\, \omega_{ab}\, t} -1\right),
 \label{xtop:eq}  \\
 \textbf{W} \!\! &=&  \!\! \frac{1}{\hbar}\, \sum_a  \frac{\partial E_a (\textbf{k})}{\partial \textbf{k}}\,Q_a,  \quad
 \textbf{Z}^{ab} =  i \, Q_a \frac{\partial Q_b}{\partial \textbf{k}},  \label{Z-op-def:eq}
\end{eqnarray}%
\end{subequations}
and $ \omega_{ab} = \frac{E_a-E_b}{\hbar}$.
Here $\textbf{W}$ is the \emph{drift velocity} and $\textbf{Z}^{ab}$ are the \emph{Zitterbewegung amplitudes}.
This is a general result for the phenomenon of the Zitterbewegung (a simple derivation of the above result and
explicit examples are presented in our recent work~\cite{Cserti_PhysRevB.81.121417}).
In what follows, these Zitterbewegung amplitudes $\textbf{Z}^{ab}$ will play a crucial role in the charge conductivity and the Chern number.

Using the Kubo formula we show that the frequency dependent charge conductivity
(often called optical conductivity) of a multi band system  can be expressed
in terms of the Zitterbewegung amplitudes $\textbf{Z}^{ab}$:
\begin{subequations}
\label{Kubo_Zitter:eq}
\begin{eqnarray}
\hspace{-4mm} \sigma_{ij}(\omega) \hspace{-2mm} &=& \hspace{-2mm} - \frac{e^2}{\hbar^2}\, \frac{1}{\omega}\,  \lim_{\delta \to 0} \textrm{Im} \Bigl\{\Pi_{ij} (\hbar \omega + i\delta)\Bigr\},   \label{sigma-def:eq}\\
\hspace{-4mm}  \Pi_{ij}(i\nu_m) \hspace{-2mm} &=& \hspace{-2mm}  
 -\frac{1}{V} \sum_{\textbf{k}} \sum_{\substack{a,b \\ a \ne b}}
 K_{ab} \, {\left(E_a  \! - \! E_b \right)}^2  \, \textrm{Tr} \Bigl[ Z^{ab}_i \, {Z_j^{ab}}^\dag \Bigr] \! , \label{Response-zitter:eq} \\
\hspace{-4mm}  K_{ab} &=& \frac{n_F(E_a -\mu) - n_F(E_b -\mu) }{i\nu_m + E_a -E_b}.   \label{K_ab-def-1:eq}
\end{eqnarray}%
\end{subequations}
Here the Zitterbewegung amplitudes $Z^{ab}_i$ ($i$ denotes the components $x,y,z$) are given by Eq.~(\ref{Z-op-def:eq}),
$\nu_m = 2\pi m/\beta$ ($m$ is an integer, $\beta = 1/(k_\textrm{B} T)$) are the bosonic Matsubara's frequencies,
$n_F(E) = 1/(e^{\beta E}+1)$ is the usual Fermi distribution, $\mu$ is the Fermi energy,
$\textrm{Im}\{\cdot \}$ is the imaginary part of the argument, $\dag$ stands for the conjugate transpose  
and finally $V$ is the volume of the sample.
The trace is taken over the band indices.
Note that $K_{ab}(z) = K_{ba}(-z) $ and $K_{aa}(z) = 0 $ for $z\ne 0$.
Equation (\ref{Kubo_Zitter:eq}) is one of our central results in this paper.
Here we omit the calculation of the Drude peak, and focus on the inter-band contribution.
One can see clearly from Eq.~(\ref{Kubo_Zitter:eq}) that the Zitterbewegung amplitudes describing the inter-band interference manifest in the inter-band contribution of the charge conductivity.
The Zitterbewegung amplitudes are indeed probed every time when one measures the conductivity of a scattering free multi band system.

\emph{Derivation of Eq.~(\ref{Kubo_Zitter:eq}).}---First consider the response function for operators $A$ and $B$
given by~\cite{Mahan_book,Bernevig:cikk,PhysRevB.74.085308,PhysRevB.78.195424}
\begin{eqnarray}\label{corrAB_def:eq}
     \widetilde{\Pi}_{AB}(i\nu_m) &=& -\frac{1}{V}\, \int_0^\beta
    \langle\, \hat{T} \, A(\tau)\, B(0)\, \rangle \, e^{i\nu_m \tau}\, d \tau \nonumber \\
  & \hspace{-20mm} = & \hspace{-12mm} \frac{1}{V\beta} \sum_{\textbf{k},n}
    \textrm{Tr} \Bigl(A \, G(\textbf{k},i\omega_n+i\nu_m)\, B\,  G(\textbf{k},i\omega_n) \Bigr),
\end{eqnarray}
where $G(\textbf{k},z)={\left[ z+ \mu -H \right]}^{-1}$ is the one-particle Green's function, $\hat{T}$ is the time ordering  operator and
$\omega_n = (2n+1)\pi/\beta$ are the fermionic Matsubara's frequencies ($n$ is an integer).
Using the projector decomposition of the Hamiltonian $H  = \sum_a E_a Q_a$
the Green's function takes the following form:
$G(\textbf{k},z)=\sum_a \frac{Q_a}{z + \mu -E_a(\textbf{k})}$.
Then substituting this Green's function into Eq.~(\ref{corrAB_def:eq}) one can find
\begin{subequations}
\label{PI-def:eq}
\begin{eqnarray}
    \widetilde{\Pi}_{AB}(i\nu_m)  &=& \!\!  \frac{1}{V}\, \sum_{\textbf{k}} \sum_{a,b} K_{ba}(i\nu_m) \,
   \textrm{Tr} \Bigl(\, A\, Q_a\, B\, Q_b\,  \Bigr), \\
   K_{ab}(i\nu_m) &=& \hspace{-2mm}
   \frac{1}{\beta}\sum_n  \frac{1}{i\omega_n + i\nu_m +\mu -E_b}\, \frac{1}{i\omega_n +\mu -E_a}
   \label{K_ab-def-2:eq}
   \nonumber \\
   &=& \hspace{-2mm} \frac{n_F(E_a -\mu) - n_F(E_b -\mu) }{i\nu_m + E_a -E_b},
\end{eqnarray}
\end{subequations}
where in the last step of the calculation of the function $K_{ab}(i\nu_m) $ we have used the usual summation technics over the Matsubara's frequencies~\cite{Mahan_book}.

Now using Eq.~(\ref{PI-def:eq}) the current-current correlation function with current operator
$\textbf{J} = \frac{\partial H}{\partial \textbf{k}}$
(in units of $e/\hbar$ which is taken into account in the expression of the conductivity) reads
\begin{eqnarray}
\label{PIJ-Ham-def:eq}
  \hspace{-5mm}  \Pi_{i j}(i\nu_m)   &\equiv&  \widetilde{\Pi}_{J_i J_j}(i\nu_m) \nonumber \\
   &=& \frac{1}{V}\, \sum_{\textbf{k}} \sum_{a,b} K_{ba} \,
   \textrm{Tr} \Bigl(\, \frac{\partial H}{\partial k_i}\, Q_a\, \frac{\partial H}{\partial k_j}\, Q_b\,  \Bigr).
\end{eqnarray}
Now using the well-known relations $Q_a \frac{\partial H}{\partial \textbf{k}} Q_b
= \delta_{ab}\, \frac{\partial E_a}{\partial \textbf{k}}\, Q_a +(E_b-E_a)\,
Q_a \frac{\partial Q_b}{\partial \textbf{k}} $  and  $Q_a Q_b = \delta_{a b} \, Q_a $, 
and the fact that $K_{aa}(z) = 0 $ for $z\ne 0$,
it is easy to obtain the current-current response function $\Pi_{ij}$ given by Eq.~(\ref{Response-zitter:eq}).
Finally, the conductivity (\ref{sigma-def:eq}) can be calculated by analytic continuation
$i \nu_m \to \hbar \omega + i \delta$, where $\delta$ is a positive infinitesimal.

\emph{The Berry curvature and the first Chern number.}---
Now, we show that the first Chern number characterizing the Hall conductivity of insulators with non-interacting Bloch electrons~\cite{Murakami_PhysRevB.69.235206,PhysRevB.78.195424}
can also be expressed in terms of the Zitterbewegung amplitudes (\ref{Z-op-def:eq}).

From  Eqs.~(\ref{PIJ-Ham-def:eq}) and~(\ref{sigma-def:eq}) we find that the intrinsic Hall conductivity
(here we focus on two dimensional systems) for $\omega \to 0 $ (dc conductivity, for $j\ne l$) is
\begin{subequations}\label{Hall_cond:eq}
\begin{eqnarray}
    \hspace{-5mm} \sigma_{jl}(\omega = 0) &=& \frac{e^2}{h}\, \frac{1}{A}\,
    \sum_\textbf{k} \sum_a  n_a \, \Omega_{jl}^{(a)}(\textbf{k}), \\
     \Omega_{jl}^{(a)}(\textbf{k})&=& 2 \pi i  \sum_{b \ne a}
    \frac{ \textrm{Tr} \left(\frac{\partial H}{\partial k_j} Q_a \frac{\partial H}{\partial k_l} Q_b \right )
    - \textrm{c.c. }}{{\left(E_a -E_b\right)}^2}  \label{TKNN_form:eq}\\
    &=& -2 \pi i \,
    \textrm{Tr} \left(Q_a \left[\frac{\partial Q_a}{\partial k_j}, \frac{\partial Q_a}{\partial k_l}\right]  \right) ,
    \label{Berry_curv_tensor:eq}
\end{eqnarray}%
\end{subequations}%
where $n_a = n_F(E_a(\textbf{k}) -\mu) $, $\Omega_{jl}^{(a)}(\textbf{k})$ is the Berry curvature (Eq.~(\ref{TKNN_form:eq}) can be casted to that derived by Thouless~\emph{et~al.}~in Ref.~\cite{TKNN_PhysRevLett.49.405}),
$A$ is the area of the sample and $c.c.$ and $[\cdot,\cdot]$~stand for the complex conjugation and the commutator, respectively.
To get (\ref{Berry_curv_tensor:eq}) we used again $Q_a \frac{\partial H}{\partial \textbf{k}} Q_b
= \delta_{ab}\, \frac{\partial E_a}{\partial \textbf{k}}\, Q_a +(E_b-E_a)\,
Q_a \frac{\partial Q_b}{\partial \textbf{k}} $ and $Q_a + \sum_{b \ne a} Q_b = I_N $.
Now replacing $Q_a$ by $Q_a = Q_a^2$ in (\ref{Berry_curv_tensor:eq}) we obtain a very simple expression for
the Berry curvature $\Omega_{jl}^{(a)}$ in terms of the Zitterbewegung amplitudes:
\begin{eqnarray}
\label{Berry_curv_tensor_2:eq}
  \Omega_{jl}^{(a)}(\textbf{k})&=& 2 \pi i \,   \textrm{Tr}\left( \left[Z_j^{aa}, {Z_l^{aa}}^\dag \right]\right). 
\end{eqnarray}%
It is interesting to note that in contrast to the Zitterbewegung,
where only the non-diagonal elements $\textbf{Z}^{ab}$ ($a \ne b$) appear in (\ref{Z-op-def:eq}),
in the Hall effect only the diagonal ones $\textbf{Z}^{aa}$ play the role.
However, in the charge conductivity for finite frequencies given by Eq.~(\ref{Kubo_Zitter:eq})
the non-diagonal elements $\textbf{Z}^{ab}$ are present.

The Hall conductivity for band insulator in which the Fermi energy $\mu$ is located
inside the energy gap between conduction and valence sub-bands and at zero temperature reads as
$\sigma_{H} = \frac{e^2}{h}\,  C_1$,
where $C_1 =  1 /(2A) \, \sum_\textbf{k} \sum_{E_a < \mu} \varepsilon_{jl}\Omega_{jl}^{(a)}(\textbf{k})$ is the first Chern number (here $\varepsilon_{jl}$ is the fully antisymmetric tensor and the summation is assumed on indices $j$ and $l$) and it can also be expressed with the Zitterbewegung amplitudes as
\begin{eqnarray}
\label{Chern_final:eq}
C_1 &=& -\frac{1}{2 \pi i} \, \int d^2 k \sum_{\substack{a \\ E_a < \mu}} \varepsilon_{jl} \textrm{Tr} \left(Z_j^{aa} {Z_l^{aa}}^\dag \right).
\end{eqnarray}%
Note that for the first Chern number one can obtained the same result starting from Eq.~(3)
given by Avron~\emph{et~al.}~in Ref.~\cite{PhysRevLett.51.51}.

\emph{An important mathematical theorem.}---Now, we recall a less known mathematical theorem
which enables us to calculate the projector operators
$Q_a = | a \rangle \langle a |$ without calculating the eigenvectors $ | a \rangle $ of the Hamiltonian $H$.
Let $H$ be an $N \times N$ hermitian matrix with $s\leq N$ distinct eigenvalues, $E_a, \dots, E_s$.
Then the matrix $H$ can be decomposed in terms of projector matrices as $H=\sum_a E_a Q_a$, where
the projector matrix $Q_a$ for $a = 1,\dots, s$ (in the mathematical literature called Frobenius covariant~\cite{Topics_in_Matrix:book}) is given by
\begin{eqnarray}
\label{Rozsa_tetel:eq}
Q_a &=& \prod_{\substack{b=1 \\ b\ne a}}^s \frac{1}{E_a-E_b}\, \left(H- E_b\, I_N \right).
\end{eqnarray}%
The proof of this theorem is based on the Cayley–-Hamilton theorem ~\cite{Topics_in_Matrix:book,Lax_Peter:book}.

\emph{Applications.}---In the following we show a few examples how the phenomenon of the Zitterbewegung is related to
the charge conductivity or the Chern number for specific multi band systems.

i)  Consider the most general non-interacting
two-band model~\cite{Bernevig:cikk,PhysRevB.78.195424} with Hamiltonian:
\begin{equation}\label{Rashba-Ham:eq}
H = \varepsilon({\bf k})\, I_2 + \textbf{h}(\textbf{k})\, \mbox{\boldmath $\sigma$},
\end{equation}
where $I_2 $ is the two by two unit matrix in spin or pseudo spin space, and the system is characterized by the one-particle energy dispersion $\varepsilon({\bf k})$ and an effective magnetic field $\textbf{h}$
depending on the wave number ${\bf k}$, while $\mbox{\boldmath $\sigma$} = (\sigma_x,\sigma_y,\sigma_z)$ is a vector formed from the Pauli matrices.
The two eigenvalues are $E_{\pm}(\textbf{k})=\varepsilon({\bf k})\pm h$, where
$h = \sqrt{{\textbf{h}}^2}$, and the two corresponding projectors obtained from Eq.~(\ref{Rozsa_tetel:eq}) are
$Q_\pm = \frac{1}{2} \left( I_2 \pm \hat{\textbf{h}}\, \mbox{\boldmath $\sigma$} \right)$,
where $\hat{\textbf{h}} = \textbf{h}/h$ is a unit vector.
The conductivity for impurity free samples can be obtained from Eq.~(\ref{Kubo_Zitter:eq}) and after a little algebra the response function becomes
\begin{eqnarray}\label{PIij_Rashba:eq}
\hspace{-2mm}   \Pi_{ij} (i\nu_m) \hspace{-2mm} &=& \hspace{-2mm} - \frac{2}{V}\,\sum_{\textbf{k}}
    \frac{n_+ - n_- }{{(i\nu_m)}^2-{(2h)}^2} \, h^2
    \nonumber \\[2ex]
& \times &
\hspace{-2mm} \left(\nu_m\, \epsilon_{\alpha \beta \gamma}
\frac{\partial \hat{h}_\alpha}{\partial k_i} \frac{\partial \hat{h}_\beta}{\partial k_j}\, \hat{h}_\gamma
+ 2 h \, \frac{\partial \hat{h}_\alpha}{\partial k_i} \frac{\partial \hat{h}_\alpha}{\partial k_j}\right),
\end{eqnarray}
where $n_\pm = n_F(E_\pm -\mu)$ and we sum on any repeated index.
For two dimensional samples this result agrees with that obtained by Bernevig~\cite{Bernevig:cikk}.

ii) For a two-band model with Hamiltonian (\ref{Rashba-Ham:eq}) using Eq.~(\ref{Chern_final:eq})
we easily find the first Chern number: $C_1 = -\frac{1}{4\pi }\, \int d^2 \textbf{k}\quad
    \hat{\textbf{h}} \cdot
    \left(
    \frac{\partial \hat{\textbf{h}}}{\partial k_x}  \times
    \frac{\partial \hat{\textbf{h}}}{\partial k_y}
    \right)$,
which is a well-known result~\cite{Murakami_PhysRevB.69.235206,PhysRevB.78.195424}.

iii) We now consider the Luttinger-type systems~\cite{Luttinger_PhysRev.102.1030,Murakami_PhysRevB.69.235206} for which the Hamiltonian is given by
\begin{equation}\label{Luttinger_Ham:eq}
    H  = \frac{\hbar^2}{2m}\, \left[ \left(\gamma_1 + \frac{5}{2}\, \gamma_2\right) \textbf{k}^2
    - 2 \gamma_2 {\left(\textbf{k} \textbf{S}\right)}^2\right],
\end{equation}
where $\textbf{k}= (k_x,k_y,k_z)$ is the wave number and $\textbf{S}= (S_x,S_y,S_z)$
represents the spin operator with spin $3/2$, while $\gamma_{1,2}$ and $m$ are parameters of the model.
The Hamiltonian can be expressed in terms of the projection operators
$Q_+$ and $Q_-$ as~\cite{Murakami_PhysRevB.69.235206}
\begin{subequations}
\label{Luttinger_Ham_proj:eq}
\begin{eqnarray}
  H &=& E_-(\textbf{k}) Q_-(\textbf{k})+E_+(\textbf{k}) Q_+(\textbf{k}), \\
  Q_+(\textbf{k}) &=& \frac{9}{8}\, I_4-\frac{1}{2\, \textbf{k}^2}\, {\left(\textbf{k} \textbf{S}\right)}^2,
  \label{QL-def:eq} \\
  Q_-(\textbf{k}) &=& I_4-Q_+(\textbf{k}),
  \label{QH-def:eq}
\end{eqnarray}%
\end{subequations}%
where $I_4$ is the $4 \times 4$ unit matrix,
and the double degenerate eigenvalues are $E_\pm(\textbf{k})= \frac{\gamma_1 \pm 2\gamma_2}{2m}\, {(\hbar\textbf{k})}^2$
corresponding to the light-hole ($+$) and the heavy-hole ($-$) bands.
The projection operators $Q_\pm$ can be obtained from Eq.~(\ref{Rozsa_tetel:eq}).
In earlier calculations of the response function $\Pi_{ij}$ the SO(5) Clifford algebra has been invoked~\cite{Murakami_PhysRevB.69.235206,Bernevig:cikk}.
In our approach the response function can be obtained without using the Clifford algebra.
Indeed, it is easy to calculate the current-current response function using
Eqs.~(\ref{Z-op-def:eq}) and~(\ref{Response-zitter:eq}) or Eq.~(\ref{PIJ-Ham-def:eq}),
and the commutation relations $\left[S_j,S_k \right]= i \, \varepsilon_{jkl}\, \hbar S_l $ for the spin operator $\textbf{S}$.
The Zitterbewegung amplitudes have already been calculated in Ref.~\cite{Cserti_PhysRevB.81.121417}.
After some algebra we have
\begin{eqnarray}\label{PIJ_Luttinger:eq}
    \hspace{-4mm} \Pi_{ij} (i\nu_m) \hspace{-2mm} &=& \hspace{-2mm} \frac{24 \hbar^6 \gamma_2^3}{m^3 {(2 \pi)}^3}\,
    \hspace{-2mm} \int d^3 k  \frac{\left(n_+ - n_- \right) \left(k^2 \delta_{ij} - k_i k_j \right)  }
    {{\left(\frac{2 \gamma_2 \hbar^2 k^2}{m}\right)}^2 - {\left(i\nu_m\right)}^2},
\end{eqnarray}
where $i,j=x,y,z$ and we used $\sum_\textbf{k} \to V \int \frac{d^3 k }{{\left(2 \pi\right)}^3}$.
The integration over the polar angles of $\textbf{k}$ can be done analytically
and at zero temperature we have the same result as that, e.g., in Ref.~\cite{Bernevig:cikk}.

iv) Consider the spin-orbit interaction in two-dimensional electron gas in a fully symmetric quantum well
investigated recently by Bernardes~\emph{et~al.}~\cite{Schliemann_Loss_ISI:000248866900038}.
The Hamiltonian of this system is given by
\begin{equation}\label{Ham_symmetric_weel:eq}
    H = \left(\begin{array}{cccc}
          \frac{\hbar^2 \textbf{k}^2}{2m} +\varepsilon_e & -i\eta k_- & 0 & 0 \\[1ex]
          i\eta k_+ & \frac{\hbar^2 \textbf{k}^2}{2m} +\varepsilon_o & 0 & 0 \\[1ex]
          0 & 0 & \frac{\hbar^2 \textbf{k}^2}{2m} +\varepsilon_o & -i\eta k_- \\[1ex]
          0 & 0 & i\eta k_+ & \frac{\hbar^2 \textbf{k}^2}{2m} +\varepsilon_e
        \end{array}
        \right),
\end{equation}
where $k_\pm = k_x \pm i k_y  $, while $\varepsilon_{e,o}$, $\eta$ and $m$ are parameters of the model.
The two double degenerate eigenvalues are
$E_\pm =\varepsilon_\textbf{k} \pm f_k$,  where
$\varepsilon_\textbf{k} = \hbar^2 \textbf{k}^2/(2m) + \varepsilon_+$,
$f_k = \sqrt{\varepsilon_-^2 + \textbf{k}^2 \eta^2}$,
$\varepsilon_\pm = (\varepsilon_e \pm \varepsilon_o)/2$.
Again one can show that $H= E_+ Q_+ + E_- Q_-$, where the two projection operators obtained from Eq.~(\ref{Rozsa_tetel:eq}) are
\begin{eqnarray}\label{Qp_Symmetric_well:eq}
   \hspace{-3mm}  Q_\pm \hspace{-2mm} &=&  \hspace{-2mm} \frac{1}{2f_k} \! \left(\begin{array}{cccc}
           h_k^\mp & \mp i\eta k_- & 0 & 0 \\[1ex]
          \pm i\eta k_+ & h_k^\pm & 0 & 0 \\[1ex]
          0 & 0 & h_k^\pm & \mp i\eta k_- \\[1ex]
          0 & 0 & \pm i\eta k_+ & h_k^\mp
        \end{array}
        \right),
\end{eqnarray}
and $h_k^\pm = \pm \varepsilon_- + f_k$.
To calculate the current-current response function we again use Eqs.~(\ref{Z-op-def:eq}) and~(\ref{Kubo_Zitter:eq}).
After some algebra we have
\begin{eqnarray}\label{PIJ_symm_well:eq}
 \hspace{-3mm} \Pi_{ij} (i\nu_m) \hspace{-2mm} &=& \hspace{-2mm} 2 \eta^2 \!\! \int \frac{d^2 k}{{\left(2\pi\right)}^2}
     \frac{\left(n_+ - n_- \right)\left(f_k^2\delta_{ij} - k_i k_j \eta^2 \right)}{f_k \left[f_k^2 -{\left(\frac{i\nu_m}{2}\right)}^2\right]},
\end{eqnarray}
where $i,j=x,y$.
Performing the integration over the polar angle of $\textbf{k}$ one can see that $ \Pi_{ij} (i\nu_m)$ is a diagonal matrix.
To our best knowledge this result is new in the literature.
The charge conductivity can be obtained from Eq.~(\ref{sigma-def:eq}) and it will be published elsewhere.

v) For single layer graphene the most general Hamiltonian in tight-binding approximation can be given by a 2 by 2 matrix in which the diagonal elements $H_{AA}$ and $H_{BB}$ include second, forth,
etc.~neighbor hopping terms,
while the off-diagonal elements $H_{AB}$ and $H_{BA}$ contain the first, third, etc.~hopping terms~\cite{Reich_PhysRevB.66.035412}.
Thus, the Hamiltonian can be mapped to Eq.~(\ref{Rashba-Ham:eq}).
Note that the same is true even for a strained graphene.
The response function can be obtained from Eq.~(\ref{PIij_Rashba:eq}).
In particular, taking into account only first-nearest neighbors, we have
$H_{AA} = H_{BB} = \varepsilon_0$ and
$H_{AB} = H_{BA}^* = f(\textbf{k})$, where
$f(\textbf{k})=\gamma_0 \left(1+e^{-i{\bf k a}_1} +e^{-i{\bf k a}_2}\right)$, and
$\gamma_0$ and $\varepsilon_0$ are parameters of the model,
and $\textbf{a}_1$ and $\textbf{a}_2$  are the unit vectors of the unit cell
in the honeycomb lattice\cite{neto:109}.
The two eigenvalues of the Hamiltonian are $E_\pm = \varepsilon_0 \pm \left|f(\textbf{k})\right| $
and $\hat{\textbf{h}} = (\textrm{Re}\{f(\textbf{k})\},-\textrm{Im}\{f(\textbf{k})\},0)/\left|f(\textbf{k})\right|$.
Using Eq.~(\ref{PIij_Rashba:eq}) the current-current response function is
\begin{subequations}\label{PIij_single-graphene:eq}
\begin{eqnarray}
    \Pi_{ij}(i\nu_m)  &=& - \frac{1}{A}\, \sum_\textbf{k}
    \frac{n_+ - n_- }{\left|f(\textbf{k})\right|^2 - {\left(\frac{i\nu_m}{2}\right)}^2 }\, F_{ij}(\textbf{k}), \\[1ex]
    F_{ij}(\textbf{k}) &=&
    \frac{\textrm{Im}\{f^* \frac{\partial f}{\partial k_i}\} \,
    \textrm{Im}\{f^* \frac{\partial f}{\partial k_j}\}}{\left|f(\textbf{k})\right|},
\end{eqnarray}%
\end{subequations}%
where the summation in $\textbf{k}$ is over the entire Brillouin zone of the honeycomb lattice,
$A$ is the area of the sample, $\textrm{Re}\{\cdot \}$ is the real part of the argument and $*$ denotes the complex conjugation.
Note that Eq.~(\ref{PIij_single-graphene:eq}) is valid not only in the usual Dirac cone approximation.
Our result agrees with that obtained by Zhang~\emph{et~al.}~\cite{PhysRevB.77.241402}
and by Yuan~\emph{et~al.}~\cite{Katsneslson_time-dep_Ham:cikk}.
However, according to our numerical calculations it slightly differs from that obtained
by Stauber~\emph{et~al.}~\cite{stauber:085432} for the high frequency region.

vi) Finally, we give at least one example in which the projector decomposition of the Hamiltonian
involves not only two projector operators (as in the previous cases) but four projector operators.
Such a system is, e.g., the bilayer graphene~\cite{Novoselov_Hall:ref,mccann:086805}.
Our general framework presented in this work for calculating the charge conductivity can also be applied to bilayer graphene.
The Zitterbewegung amplitudes have already been given in Ref.~\cite{Cserti_PhysRevB.81.121417} for bilayer graphene excluding trigonal warping.
Using this result we obtained the same analytical expression for the frequency dependent optical conductivity as that by Nicol and Carbotte in Ref.~\cite{PhysRevB.77.155409} using the spectral function representation of the Green's function.
We would like to stress that our approach for calculating the optical conductivity is a convenient and very efficient method even for more complex systems.
The study of strained bilayer graphene with/without trigonal warping is in progress.

\emph{Conclusions.}---In this paper we derived an explicit expression for the charge conductivity, the Berry curvature  and the Chern number in terms of the Zitterbewegung amplitudes.
Our results show that the Zitterbewegung is not just an interesting phenomenon
in quantum physics but it is closely related to measurable effect.

We would like to thank B. A. Bernevig, C. W. J. Beenakker, B. B\'eri, M. I. Katsnelson, L. Oroszl\'any
N. M. R. Peres, and A. P\'alyi for helpful discussions.
This work is supported by the Marie Curie ITN project NanoCTM (FP7-PEOPLE-ITN-2008-234970)
and the Hungarian Science Foundation OTKA under the contracts No. 75529 and No.~81492.


\begin{thebibliography}{10}


\bibitem{Schrodinger:cikk}
E. Schr\"odinger, Sitzungsber. Preuss. Akad. Wiss. Phys. Math. Kl. {\bf 24},
  418  (1930).

\bibitem{Schliemann_Loss_PhysRevLett.94.206801}
J. Schliemann, D. Loss, and R.~M. Westervelt, Phys. Rev. Lett. {\bf 94},
  206801  (2005).

\bibitem{Cserti_Zitter-ISI:000242409000014}
J. Cserti and G. David, Phys. Rev. B {\bf 74},  172305  (2006).

\bibitem{Cserti_PhysRevB.81.121417}
G. D\'avid and J. Cserti, Phys. Rev. B {\bf 81},  121417 (R)  (2010).
For an extended version see arXiv:0909.2004v3.

\bibitem{Katsnelson_Zitter_min-cond:cikk}
M.~I. Katsnelson, Eur. Phys. J. B {\bf 51},  157  (2006).

\bibitem{Murakami_PhysRevB.69.235206}
S. Murakami, N. Nagaosa, and S.-C. Zhang, Phys. Rev. B {\bf 69},  235206
  (2004).

\bibitem{PhysRevB.78.195424}
X.-L. Qi, T.~L. Hughes, and S.-C. Zhang, Phys. Rev. B {\bf 78},  195424
  (2008).

\bibitem{Mahan_book}
G.~D. Mahan, {\em Many-Particle Physics} (Plenum Press, 2nd ed., New York and
  London, 1990).

\bibitem{Bernevig:cikk}
B.~A. Bernevig, Phys. Rev. B {\bf 71},  073201  (2005).

\bibitem{PhysRevB.74.085308}
X.-L. Qi, Y.-S. Wu, and S.-C. Zhang, Phys. Rev. B {\bf 74},  085308  (2006).

\bibitem{TKNN_PhysRevLett.49.405}
D.~J. Thouless, M. Kohmoto, M.~P. Nightingale, and M. den Nijs, Phys. Rev.
  Lett. {\bf 49},  405  (1982).

\bibitem{PhysRevLett.51.51}
J.~E. Avron, R. Seiler, and B. Simon, Phys. Rev. Lett. {\bf 51},  51  (1983).

\bibitem{Topics_in_Matrix:book}
R.~A. Horn and C.~R. Johnson, {\em Topics in Matrix Analysis} (Cambridge
  University Press, Cambridge, UK, 1991).

\bibitem{Lax_Peter:book}
P.~D. Lax, {\em Linear Algebra and Its Applications} (John Wiley and Sons Inc.,
  2nd ed., New York, United States, 2007).

\bibitem{Luttinger_PhysRev.102.1030}
J.~M. Luttinger, Phys. Rev. {\bf 102},  1030  (1956).

\bibitem{Schliemann_Loss_ISI:000248866900038}
E. Bernardes {\it et~al.}, Phys. Rev. Lett. {\bf 99},  076603  (2007).

\bibitem{Reich_PhysRevB.66.035412}
S. Reich, J. Maultzsch, C. Thomsen, and P. Ordej\'on, Phys. Rev. B {\bf 66},
  035412  (2002).

\bibitem{neto:109}
A.~H.~C. Neto {\it et~al.}, Rev. Mod. Phys. {\bf 81},  109  (2009).

\bibitem{PhysRevB.77.241402}
C. Zhang, L. Chen, and Z. Ma, Phys. Rev. B {\bf 77},  241402  (2008).

\bibitem{Katsneslson_time-dep_Ham:cikk}
S. Yuan, H.~D. Raedt, and M.~I. Katsnelson, unpublished, arXiv:1007.3930v1.

\bibitem{stauber:085432}
T. Stauber, N.~M.~R. Peres, and A.~K. Geim, Phys. Rev. B {\bf 78},  085432
  (2008).

\bibitem{Novoselov_Hall:ref}
K. Novoselov {\it et~al.}, Nature Phys. {\bf {2}},  177  ({2006}).

\bibitem{mccann:086805}
E. McCann and V.~I. Fal'ko, Phys. Rev. Lett. {\bf 96},  086805  (2006).

\bibitem{PhysRevB.77.155409}
E.~J. Nicol and J.~P. Carbotte, Phys. Rev. B {\bf 77},  155409  (2008).

\end{thebibliography}

\end{document}